# Unconventional superconductivity in ζ-phase of oxygen compressed to megabar pressure

E. F. Talantsev[1,2]

[1]M.N. Miheev Institute of Metal Physics, Ural Branch, Russian Academy of Sciences, 18, S. Kovalevskoy St., Ekaterinburg, 620108, Russia

[2]NANOTECH Centre, Ural Federal University, 19 Mira St., Ekaterinburg, 620002, Russia

E-mail: evgeny.talantsev@imp.uran.ru

## Abstract

Oxygen exhibits structural phase transformation from non-metallic ε-$O_2$ phase into metallic ζ-$O_2$ phase at pressure of $P$ = 96 GPa (Desgreniers *et al* 1990 *J. Phys. Chem.* **94** 1117; Akahama *et al* 1995 *Phys. Rev. Lett.* **74** 4690). Metallic ζ-$O_2$ phase is a superconductor with transition temperature of $T_c$ = 0.6 K at $P$ = 115-120 GPa (Shimizu *et al* 1998 *Nature* **393** 767). In this paper we have performed analysis of temperature dependent upper critical field, $B_{c2}(T)$, for ζ-$O_2$ phase ($P$ = 115 GPa) and show that this highly-compressed phase of gaseous molecular element is unconventional superconductor with the ratio of $T_c$ to the Fermi temperature, $T_F$, in the range of $0.009 \leq T_c/T_F \leq 0.108$.

## Unconventional superconductivity in ζ-phase of oxygen compressed to megabar pressure

## I. Introduction

Jörg Wittig [1] heralded studies of pressure-induced superconductivity in non-superconductors by the discovery of superconducting transition in cerium with $T_c$ = 1.8 K by applying isostatic pressure of $P$ = 5 GPa. To date, the pressure-induced superconductivity has been detected in dozens of elements and compounds compressed at megabar pressures [2-18], including milestone experimental discoveries of near-room-temperature (NRT) superconducting hydrides [2-4,10-13].

Desgreniers *et al* [19] detected metallization of oxygen at pressures $P$ > 90 GPa, and this phase transformation from non-metallic ε-$O_2$ phase into metallic ζ-$O_2$ phase has been studied for three decades [20-23].

Shimizu *et al* [8] reported that ζ-$O_2$ phase is a superconductor with transition temperature of $T_c$ = 0.6 K at pressures in the range of $P$ = 115-120 GPa. Shimizu *et al* [8] also reported temperature dependent upper critical field data, $B_{c2}(T)$, which we analyse herein with the purpose to classify superconductivity (i.e., conventional vs unconventional) in ζ-$O_2$ phase.

Primary demand to perform this classification is based on our recent findings that all NRT highly-compressed superconductors are, surprisingly enough, unconventional superconductors [24-26], which exhibit the ratio of the superconducting transition temperature, $T_c$, to the Fermi temperature, $T_F$, in the range of $0.01 \lesssim \frac{T_c}{T_F} \lesssim 0.05$, which is the range for all known unconventional superconductors [27,28].

In result, we find that in all considered scenarios ζ-$O_2$ phase ($P$ = 115 GPa) has $T_c/T_F$ ratio in the range of $0.008 \lesssim \frac{T_c}{T_F} \lesssim 0.107$, and, thus, this highly-compressed phase of gaseous element should be classified as unconventional superconductor.

## II. The upper critical field data analysis

Shimizu *et al* [8] in their Fig. 3 reported experimental $B_{c2}(T)$ data for ζ-$O_2$ phase at pressure $P$ = 115 GPa which we fit to two models:

1. The first model was proposed by Baumgartner *et al* [29]:

$$B_{c2}(T) = \frac{\phi_0}{2\cdot\pi\cdot\xi^2(0)} \cdot \left( \frac{\left(1-\frac{T}{T_c}\right) - 0.153\cdot\left(1-\frac{T}{T_c}\right)^2 - 0.152\cdot\left(1-\frac{T}{T_c}\right)^4}{0.693} \right) \tag{1}$$

where $\phi_0 = 2.068 \cdot 10^{-15}$ Wb is magnetic flux quantum, and ξ(0) is the ground state coherence length. We will designate this model as B-WHH model, because Eq.1 is analytical approximation of general model proposed by Werthamer, Helfand, and Hohenberg [30].

2. The second used model was proposed by Gor'kov [31]:

$$B_{c2}(T) = \frac{\phi_0}{2\cdot\pi\cdot\xi^2(0)} \cdot \left( \frac{1.77 - 0.43\cdot\left(\frac{T}{T_c}\right)^2 + 0.07\cdot\left(\frac{T}{T_c}\right)^4}{1.77} \right) \cdot \left[1 - \left(\frac{T}{T_c}\right)^2\right] \tag{2}$$

It should be noted, that both models (Eqs. 1,2) are in wide use at the moment to analyse experimental $B_{c2}(T)$ data for whole variety of superconducting materials, ranging from atomically thin superconductors [32] and practical superconductors [29] to NRT superconductors [24-26].

Results of fit of $B_{c2}(T)$ data for ζ-$O_2$ phase ($P$ = 115 GPa) to Eqs. 1,2 are shown in Fig. 1 and deduced parameters are collected in Table 1.

**Table I.** Deduced and calculated parameters for ζ-$O_2$ phase compressed at pressure of $P$ = 115 GPa. The smallest and the largest values for $\frac{T_c}{T_F}$ are marked in bold.

| Model | Deduced $T_c$ (K) | Deduced ξ(0) (nm) | Assumed $m^*_{eff}$ (in $m_e$) | Assumed $\frac{2\cdot\Delta(0)}{k_B\cdot T_c}$ | $T_F$ (K) | $T_c/T_F$ |
|---|---|---|---|---|---|---|
| B-WHH | 0.635 ± 0.008 | 41.3 ± 0.4 | 0.49 | 3.53 | 5.86 ± 0.02 | **0.108 ± 0.003** |
| | | | | 5.0 | 11.8 ± 0.02 | 0.052 ± 0.002 |
| | | | 3.0 | 3.53 | 35.9 ± 1.6 | 0.018 ± 0.001 |
| | | | | 5.0 | 72.0 ± 3.2 | **0.009 ± 0.001** |
| Gor'kov | 0.63 ± 0.01 | 42.0 ± 0.7 | 0.49 | 3.53 | 5.96 ± 0.40 | 0.106 ± 0.005 |
| | | | | 5.0 | 12.0 ± 0.02 | 0.053 ± 0.003 |
| | | | 3.0 | 3.53 | 36.5 ± 2.5 | 0.017 ± 0.001 |
| | | | | 5.0 | 73.3 ± 4.8 | 0.009 ± 0.001 |

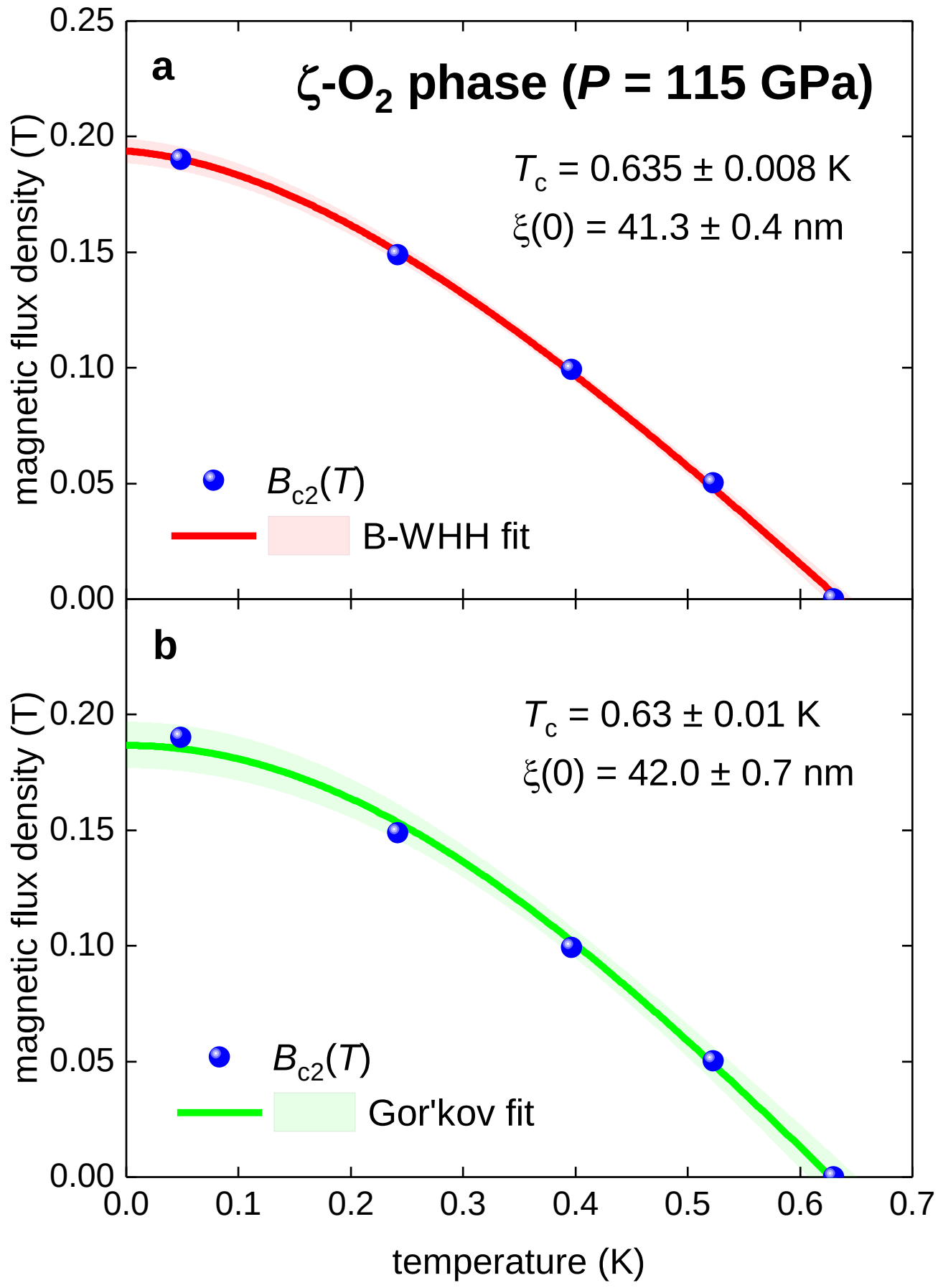

**Figure 1.** Superconducting upper critical field, $B_{c2}(T)$, data and fits to three models (Eqs. 1,2) for ζ-$O_2$ phase compressed at pressure of $P$ = 115 GPa (raw data is from Ref. 8). (a) fit to B-WHH model, $R$ = 0.9994; (b) fit to Gor'kov model, $R$ = 0.998. 95% confidence bars are shown.

## III. ζ-$O_2$ phase in Uemura plot

From known ξ(0) and $T_c$ values, the Fermi temperature, $T_F$, can be calculated by an equation of the Bardeen-Cooper-Schrieffer (BCS) theory [33]:

$$T_F = \frac{\varepsilon_F}{k_B} = \frac{\pi^2}{8} \cdot m_{eff}^* \cdot \xi^2(0) \cdot \left(\frac{\alpha \cdot k_B \cdot T_c}{\hbar}\right)^2, \quad (3)$$

where $\alpha = \frac{2 \cdot \Delta(0)}{k_B \cdot T_c}$, Δ(0) is the amplitude of the ground state energy gap, $\varepsilon_F$ is the Fermi energy, $\hbar = h/2\pi$ is reduced Planck constant, $k_B$ is the Boltzmann constant, $m_{eff}^*$ is the charge carrier effective mass.

As there are no available experimental $\alpha = \frac{2 \cdot \Delta(0)}{k_B \cdot T_c}$ and the effective charge carrier mass $m_{eff}^*$ values for ζ-$O_2$ phase, to calculate $T_F$ for ζ-$O_2$ phase we chose a reasonable lower and upper bounds for these values. For lower bound of $m_{eff}^*$ we use the value for ambient pressure hydrogen-rich superconductor, $PdH_x$ [34]:

$$m_{eff}^* = 0.49 \cdot m_e, \quad (4)$$

despite a fact that the closest (by $T_c$ and by the atomic mass) ambient pressure superconductor to ζ-$O_2$ is the aluminium [35] with

$$m_{eff}^* = 1.0 \cdot m_e \quad (5)$$

In this regard, in lower bound for $m_{eff}^*$ (Eq. 4) is chosen as intendent underestimated value to cover some hypothetical case that $m_{eff}^*$ might be reasonably low in ζ-$O_2$ phase. This lower bound for $m_{eff}^*$ can be also supported by $m_{eff}^* = (0.2 - 0.5) \cdot m_e$ reported by Medvedeva [36] for multicomponent conducting oxides.

For the upper bound of $m_{eff}^*$ we use the highest value reported for highly compressed hydrides, $m_{eff}^* = 3.0 \cdot m_e$ [37]. The possibility for heavy effective charge carrier mass in ζ-$O_2$, i.e. $m_{eff}^* > 3.0 \cdot m_e$, cannot be of course rejected a priori, but large effective masses, as a rule, always associated with strong interaction between spin and *d*- or *f*-orbitals, and the latter does not exist in such light elements, like oxygen. Thus, we do not consider a possibility that the effective mass can exceed mentioned above value of $m_{eff}^* = 3.0 \cdot m_e$.

The lowest value for $\frac{2 \cdot \Delta(0)}{k_B \cdot T_c}$ is weak-coupling limit of 3.53 [33], and for all known *s*-wave superconductors [38-40] α is limited by the upper bound of $\frac{2 \cdot \Delta(0)}{k_B \cdot T_c} \lesssim 5.0$ [38]. Thus, $T_F$ is calculated (Table 1 and Fig. 2) in the assumption that α is varying within a range of $3.53 \leq \frac{2 \cdot \Delta(0)}{k_B \cdot T_c} \leq 5.0$ (it should be noted, that this range is cover most highly-compressed hydrogen-rich superconductors [24,37-42]).

As the result, ζ-$O_2$ phase ($P$ = 115 GPa) in all considered scenarios (Table 1) has 0.009 ≤ $T_c/T_F$ ≤ 0.108 and falls in unconventional superconductors band of the Uemura plot [27,28] (Fig. 2).

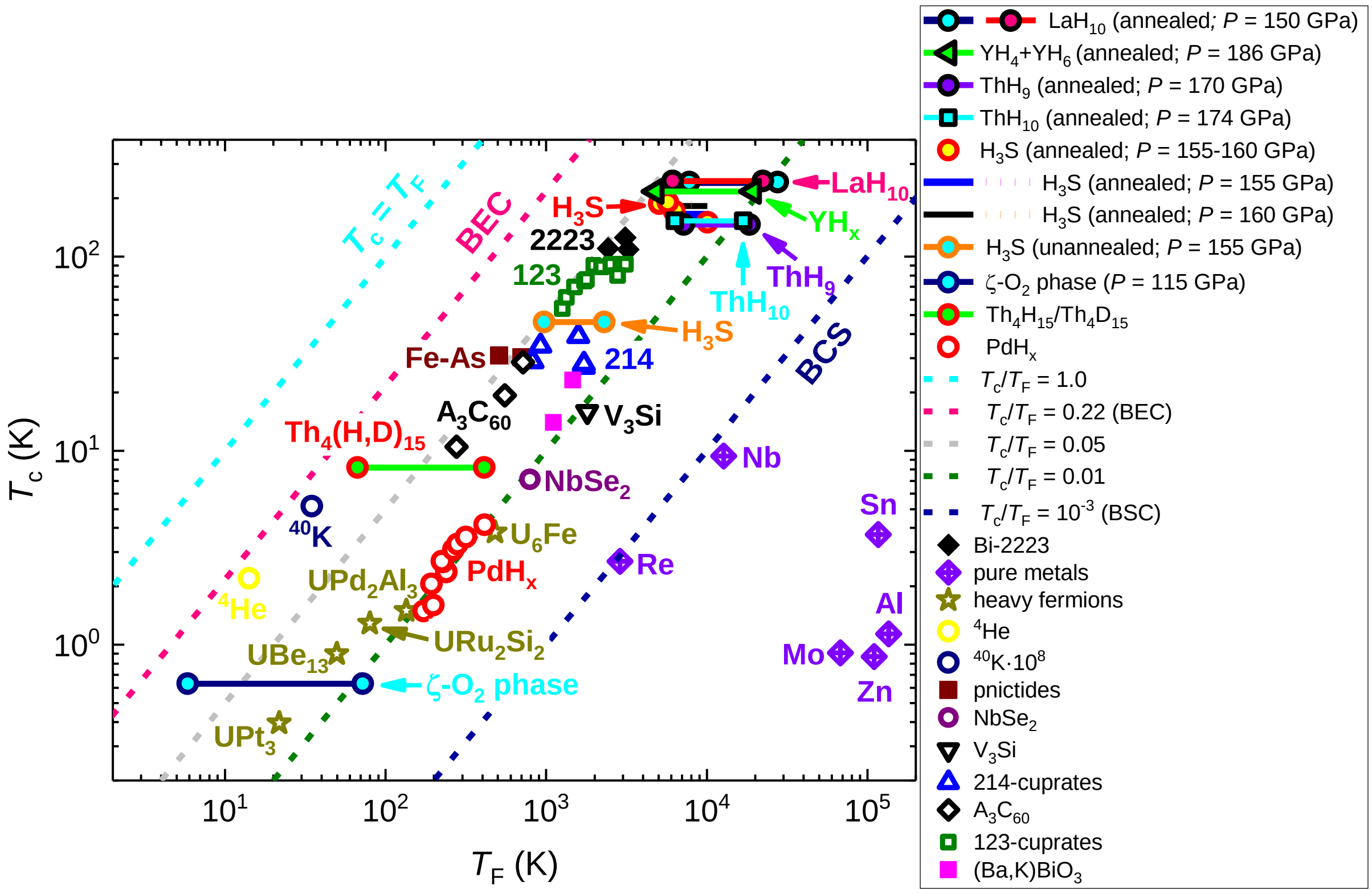

**Figure 2.** A plot of $T_c$ versus $T_F$ where ζ-$O_2$ phase compressed at pressure of $P$ = 115 GPa is shown together with the most representative superconducting families. Raw data is taken from [24-28]. Characteristic lines for the Bose-Einstein condensate (BEC), the Bardeen-Cooper-Schrieffer (BCS) superconductors and for $T_c/T_F$ = 1.0, 0.05, 0.01 are shown for clarity.

It should be stressed that primary physical reason, that ζ-$O_2$ phase is classified as unconventional superconductor is belong solid experimental result, that this superconductor exhibits relatively high ground state upper critical field, $B_{c2}(0)$, and relatively low superconducting transition temperature, $T_c$. Truly, if ζ-$O_2$ phase will be conventional superconductor similar to Al (which exhibits $T_c/T_F = 8.4 \cdot 10^{-6}$), than in accordance with the expression [25] based on BCS theory [33]:

$$B_{c2}\left(\frac{T}{T_c} = 0\right) = \frac{\pi \cdot \phi_0 \cdot k_B}{16 \cdot \hbar^2} \cdot m_{eff}^{*} \cdot \alpha^2 \cdot \left(\frac{T_c}{T_F}\right) \cdot T_c = 2.7\ mT \quad (6)$$

where for the simplicity we assumed that $m_{eff}^{*} = 1.0 \cdot m_e$ and $\alpha = \frac{2 \cdot \Delta(0)}{k_B \cdot T_c} = 3.53$ are identical to ones for aluminium. We note, that experimental value for $\zeta$-$O_2$ phase [8] is:

$$B_{c2}\left(\frac{T}{T_c} = 0.08\right) = 190\ mT, \quad (7)$$

which is about two orders of magnitude larger than hypothetical value for a conventional superconductor.

## V. Conclusions

In summary, in this paper we analyse experimental $B_{c2}(T)$ data for superconducting $\zeta$-$O_2$ phase of highly-compressed oxygen ($P$ = 115 GPa), reported by Shimizu *et al* [8], and find that in all considered scenarios this highly-compressed $\zeta$-$O_2$ phase is unconventional superconductor.

**Acknowledgement**

Author also thanks financial support provided by the state assignment of Minobrnauki of Russia (theme “Pressure” No. AAAA-A18-118020190104-3) and by Act 211 Government of the Russian Federation, contract No. 02.A03.21.0006.